\documentclass[letterpaper, 10 pt, conference]{ieeeconf}  

\IEEEoverridecommandlockouts                              

\usepackage{graphicx} 
\usepackage{epsfig} 
\usepackage{amsmath} 
\usepackage{amssymb}  
\usepackage{mathtools}
\usepackage{algorithm}
\usepackage{algpseudocode}
\usepackage{tikz}
\usetikzlibrary{fit}
\usetikzlibrary{arrows} 
\usetikzlibrary{decorations.pathmorphing}
\usetikzlibrary{matrix} 
\usetikzlibrary{calc}
\usetikzlibrary{backgrounds}

\usepackage{xcolor}
\usepackage{hyperref}
\usepackage[noadjust]{cite}

\newtheorem{theorem}{Theorem}
\newtheorem{proposition}{Proposition}
\newtheorem{definition}{Definition}
\newtheorem{assumption}{Assumption}
\newtheorem{remark}{Remark}

\newcommand{\KCBF}{\mathcal{U}_{\mathrm{CBF}}}
\newcommand{\hpb}{h_{\mathrm{PB}}}
\newcommand{\ahpb}{\hat{h}_{\mathrm{PB}}}
\newcommand{\approxhpb}{\hat{h}_{\mathrm{PB}}}
\newcommand{\Spb}{\mathcal{S}_{\mathrm{PB}}}
\newcommand{\Dpb}{\mathcal{D}_{\mathrm{PB}}}

\title{\LARGE \bf
Approximate Predictive Control Barrier Functions using Neural Networks: A Computationally Cheap and Permissive Safety Filter
}

\author{Alexandre Didier$^{*,1}$, Robin C. Jacobs$^{*,1}$, Jerome Sieber$^{1}$, Kim P. Wabersich$^{2}$, and Melanie N. Zeilinger$^{1}$
\thanks{$^*$ Alexandre Didier and Robin C. Jacobs contributed equally to this paper.}%
\thanks{$^{1}$ are members of the Institute for Dynamic Systems and Control, ETH Zurich, Zurich 8092, Switzerland. \newline {\tt\small \{adidier,jacobsr,jsieber,mzeilinger\}@ethz.ch}}%
\thanks{$^{2}$ is a member of the Corporate Research of Robert Bosch GmbH, Renningen, 71272, Germany. {\tt\small wabersich@kimpeter.de} }%
}

\begin{document}

\maketitle
\thispagestyle{empty}
\pagestyle{empty}

\begin{abstract}
A predictive control barrier function (PCBF) based safety filter is a modular framework to verify safety of a control input by predicting a future trajectory. The approach relies on the solution of two optimization problems, first computing the minimal state constraint violation given the current state in the form of slacks on the constraint, and then computing the minimal deviation from a proposed input given the previously computed minimal slacks. This paper presents an approximation procedure that uses a neural network to approximate the optimal value function of the first optimization problem, which defines a control barrier function (CBF). By including this explicit approximation in a CBF-based safety filter formulation, the online computation becomes independent of the prediction horizon. It is shown that this approximation guarantees convergence to a neighborhood of the feasible set of the PCBF safety filter problem with zero constraint violation.
The convergence result relies on a novel class $\mathcal{K}$ lower bound on the PCBF decrease and depends on the approximation error of the neural network. Lastly, we demonstrate our approach in simulation for an autonomous driving example and show that the proposed approximation leads to a significant decrease in computation time compared to the original approach.
\end{abstract}

\section{INTRODUCTION }
A main limitation of learning-based control algorithms, which have demonstrated wide-spread success in various applications such as, e.g., reinforcement learning in \cite{bertsekas2019reinforcement} for highly dynamic locomotion of robotic systems, see \cite{hwangbo19, haarnoja}, is their lack of rigorous safety guarantees.
As a result, there has been a growing interest in modular safety frameworks, which allow for verifying safety of control inputs.

One example of such a modular framework is the predictive safety filter \cite{wabersich18}, which is able to provide constraint satisfaction guarantees by solving an online optimization problem, whose constraints are based on a model predictive control (MPC) formulation, see \cite{rawlings_model_2017} for an overview.
At every time step, the closest safe input to the proposed control input is computed, such that an invariant terminal set can be reached by a model-based trajectory prediction, all while satisfying state and input constraints. 
While predictive safety filter methods can also be designed to fulfill constraints robustly for bounded uncertainties or with a certain probability in the stochastic case, see, e.g., \cite{wabersich18} and \cite{wabersich21}, respectively, infeasible initial conditions or inaccurate uncertainty descriptions compromise their theoretical guarantees.
To remedy these issues, slack variables can be introduced which allow for a relaxation of the state constraints, as done e.g., in~\cite{tearle21} for a miniature race car example. However, the introduction of slack variables leads to a loss of rigorous safety guarantees.
Recent results in~\cite{wabersich22} enable the use of soft state constraints, while guaranteeing that the system will return into the feasible set of the original safety filter problem.
The idea is to move the computation of the minimal slacks into a separate optimization problem with increasingly tightened state constraints and a terminal safe set defined by a control barrier function (CBF), see, e.g., \cite{ames2019} for an overview. The optimal slacks are then used in the safety filter problem. The safety guarantees follow from showing that the optimal value function of the slack optimization problem is a CBF. While this scheme combines the beneficial properties of CBFs and predictive safety filters, it requires the solution of two optimization problems online, where in addition, long planning horizons can potentially be required to reduce conservativeness of the safety filter. The increased computational requirements can render the approach unsuitable for embedded systems with stringent limitations on the computation time.

\textit{Contributions:}
This paper proposes a safety filter based on the approximation of the optimal value function of the slack computation, denoted predictive control barrier function~(PCBF), which was introduced in~\cite{wabersich22}.
The approximation is based on a neural network (NN), whose explicit formulation is then used in a CBF-based safety filter formulation see, e.g., \cite{ames2019}, rendering the online computation time independent of the planning horizon.
We provide a novel, strictly increasing lower bound of a continuous PCBF decrease function as introduced in \cite{wabersich22}, which is a positive definite function with respect to the feasible set of the PCBF safety filter with zero constraint violation and which bounds the decrease of the PCBF at every time step.
This bound enables showing that convergence to the feasible set of the original predictive safety filter problem is preserved for a sufficiently small NN approximation error. 
Furthermore, we demonstrate the proposed method in simulation for an autonomous driving example with different practical design choices.

\textit{Related Work:}
Numerous works approximate an MPC policy with a neural network, such as \cite{ hertneck2018, paulson20, nubert20, chen22}, while guaranteeing stability and robust constraint satisfaction, e.g., through robustness with respect to input disturbances. However, the policy of the PCBF-based safety filter is generally not continuous and therefore not suitable for a neural network approximation. 
Learning the optimal value function of an MPC controller is considered, e.g., in \cite{tamar2017learning} and \cite{karnchanachari2020practical}, to improve controller performance.
In this work, the optimal value function of an MPC-based optimization problem, which fulfills the definition of a CBF, for discrete-time continuous systems is approximated offline and directly integrated in a safety filter formulation.
An overview of methods for safety certificate learning with neural networks is provided primarily focusing on continuous-time systems in \cite{dawson2023safe}. The techniques could be used for discrete-time continuous dynamics considered in this work as a means to obtain a terminal CBF, which can then be further enhanced with the proposed method. 
Methods which use learning specifically for CBFs are, e.g., considered in \cite{robey20}, where a support vector machine is used and in \cite{srinivasan20} where online sensor data is used to learn a CBF for Lipschitz continuous, continuous-time dynamics from expert demonstrations.
In \cite{taylor20}, the time derivative of a continuous-time CBF is learned to episodically decrease model uncertainty of a safety filter method. 
Another approach to construct a safety filter is based on Hamilton-Jacobi (HJ) reachability analysis, such as, e.g., \cite{choi2021robust}, which establishes a link of HJ reachability analysis to CBFs for disturbed systems, and which can be solved in an approximate dynamic programming fashion using deep learning as done in \cite{bansal2021deepreach}.



\section{PRELIMINARIES}
\label{sec:preliminaries}
We consider discrete-time dynamical systems described by a continuous function $f: \mathbb{R}^n \times \mathbb{R}^m \to \mathbb{R}^n$ such that
\begin{equation}
\label{eq:sys}
x(k+1) = f(x(k),u(k)) \;\; \forall k \in \mathbb{N},\\
\end{equation}
where we denote the state at time step $k$ as $x(k) \in \mathbb{R}^n$ and the input applied to the system as $u(k) \in \mathbb{R}^m$. The initial state is given by $x(0) = x_0$. 
At each time step $k\in \mathbb{N}$, system \eqref{eq:sys} is subject to state constraints of the form $
x(k) \in \mathcal{X} := \{x \in \mathbb{R}^n \mid c_x(x) \leq 0\}$ and input constraints  $u(k) \in \mathcal{U} := \{u \in \mathbb{R}^{m}  \mid c_u(u) \leq 0\}$,   where $c_x : \mathbb{R}^n \to \mathbb{R}^{n_x}$ 
 and $c_u : \mathbb{R}^m \to \mathbb{R}^{n_u}$ are continuous. Both the state as well as the input constraint sets $\mathcal{X}$ and $\mathcal{U}$ are assumed to be compact. 

The aim is to apply inputs $u_p(k) \in \mathbb{R}^m$ proposed by a task-specific controller, e.g., a policy learned via reinforcement learning or a human input, to system \eqref{eq:sys} while ensuring safety of the system. In order to guarantee safety at all time steps, inputs that lead to future constraint violations need to be detected and modified online. A set of states is therefore deemed safe if an input exists for which state and input constraint satisfaction can be guaranteed for all future time steps within this set.
\begin{definition}[Safe Set]\label{def:safeset}
 A set $\mathcal{S} \subseteq \mathcal{X} $ is called a safe set, if $\forall x \in \mathcal{S}$ there exists an input $u\in \mathcal{U}$ such that $f(x,u) \in \mathcal{S}$.
\end{definition}

In order to verify safety of inputs $u_p(k)$, we consider predictive safety filters as first proposed in \cite{wabersich18}. The predictive safety filter computes a system trajectory, fulfilling state and input constraints,
to a terminal safe set in an online optimization problem and thereby approximates the maximal safe set.
This formulation considered in this work includes relaxed state and terminal constraints, as proposed in \cite{wabersich22}, to account for potentially infeasible initial conditions or unexpected disturbances and is given by
\begin{subequations}
\label{eq:pcbf_sf}
\begin{align}
 \{ u^*_{i|k} \} \in \arg \min_{u_{i|k}} \; \; & \lvert \lvert{u_p(k) - u_{0|k}}\rvert \rvert \\
\text{s.t.}\; \; &\forall\, i = 0,\dots,N-1, \nonumber \\
& x_{0|k} = x(k), \\
& x_{i+1|k} = f(x_{i|k}, u_{i|k}), \\
& u_{i|k} \in \mathcal{U}, \label{eq:pcbf_inputconstraints}\\
& x_{i|k} \in \overline{\mathcal{X}}_i( \xi^*_{i|k}), \label{eq:pcbf_sf_stateconstraints} \\
&h_f(x_{N|k}) \leq  \xi^*_{N|k}, \label{eq:pcbf_sf_terminalconstraints}
\end{align}
\end{subequations}

The $i$-step-ahead predicted state and input variables at time $k$ are denoted as $x_{i|k}$ and $u_{i|k}$ and the planning horizon as $N$, respectively.
The predicted states are subject to 
relaxed state constraints via the minimal slacks $\xi_{i|k}^*\geq0$, the exact definition of which is provided in the following in \eqref{eq:pcbf}. The constraints in \eqref{eq:pcbf_sf_stateconstraints} are additionally subject to a progressive tightening, i.e., $\overline{\mathcal{X}}_i(\xi) := \{ x\in \mathbb{R}^n \mid c_x(x) \leq -\Delta_i \mathbf{1} + \xi  \},$ where strictly increasing tightenings $\Delta_i \in \mathbb{R}_{\geq0}, \Delta_i<\Delta_{i+1} $ are used with $\Delta_0=0$ and $\mathbf{1}$ denoting the vector of ones. The strictly increasing tightenings allow for establishing the main result in \cite{wabersich22}, which is provided in Theorem~\ref{thm:1}.
Note that the input constraints $\mathcal{U}$ are not relaxed in \eqref{eq:pcbf_inputconstraints} as they typically represent physical limitations of the actuators. 
The terminal constraint \eqref{eq:pcbf_sf_terminalconstraints} is given with respect to a CBF $h_f$, which fulfills the following definition in \cite[Definition III.1]{wabersich22}.
\begin{definition}[Control barrier function]
\label{def:CBF}
A function $h: \mathcal{D} \to \mathbb{R}$ is a discrete-time control barrier function with safe set $\mathcal{S} = \{x \in \mathbb{R}^n \mid h(x) \leq 0 \} \subset \mathcal{D}$
 if the following hold:\begin{enumerate}
    \item[(1)] $\mathcal{S}, \mathcal{D}  $ are compact and non-empty sets,
    \item[(2)] $h(x)$ is continuous on $\mathcal{D}$,
    \item[(3)] there exists a continuous decrease function $\Delta h : \mathcal{D} \to \mathbb{R}$ with $\Delta h(x){>}0 $ for all $x \in \mathcal{D} \setminus \mathcal{S} $ such that:
    \begin{subequations}
    \label{eq:rate_cond_cbf}
    \begin{align}
    \hspace{-0.2cm}
        \forall x \in \mathcal{D} \setminus \mathcal{S} \: :& \inf_{u\in \mathcal{U}} h(f(x,u))- h(x) \leq - \Delta h(x) \label{eq:outside_cbf_condition},\\
        \forall x \in \mathcal{S} \: :& \inf_{u\in \mathcal{U}}h(f(x,u)) \leq 0. \label{eq:inside_cbf_condition}
    \end{align}
    \end{subequations}
\end{enumerate}
Given a CBF $h$, an input $u$ applied to system \eqref{eq:sys} at state $x$ is said to be \textit{safe}, if $u \in \KCBF(x) \subseteq \mathcal{U}$ with
\begin{align} \label{eq:cbf_ctrl_law_cond}
    \KCBF(x) \coloneqq \left\{ \begin{array}{ll} 
    U_1(x), & \text{if } x\in \mathcal{D} \setminus \mathcal{S},\\
    U_2(x), & \text{if } x\in \mathcal{S},
    \end{array}\right.
\end{align}
where $U_1(x) \coloneqq \{ u \in \mathcal{U} \mid h(f(x,u)) - h(x) \leq -\Delta h(x)\}$ and $U_2(x) \coloneqq \{ u \in \mathcal{U} \mid h(f(x,u)) \leq 0 \}$.
\end{definition}

The terminal constraint \eqref{eq:pcbf_sf_terminalconstraints} therefore requires the predicted state $x_{N|k}$ to lie inside the safe set defined as the sublevel set of the CBF $h_f$, i.e., $\mathcal{S}_f := \{ x \in \mathcal{X} \mid h_f(x) \leq 0 \}$, albeit relaxed with the terminal slack $\xi^*_{N|k}$. The terminal safe set $\mathcal{S}_f$ is required to lie within the tightened state constraints $\mathcal{S}_f\subset\overline{\mathcal{X}}_{N-1}(0)$ and the domain of $h_f$ is defined as $\mathcal{D}_{f} : = \{ x \in \mathbb{R}^n \mid h_f(x) \leq \gamma_f \}$, where $\gamma_f > 0$. Details on how to design such a CBF are provided in \cite[Section IV]{wabersich22}. We note that formulation~\eqref{eq:pcbf_sf} reduces to a nominal predictive safety filter if $\Delta_i=0$ and $\xi^*_{i|k}=0$ and can be modified to ensure robustness properties similar to \cite{wabersich18}. 

\begin{figure}
    \centering
    \vspace{0.1cm}


\tikzstyle{block} = [draw,rectangle,thick,minimum height=2.5em,minimum width=5em]
\tikzstyle{blockfixwidth} = [draw,rectangle,thick, minimum height=2.5em, minimum width=13em]
\tikzstyle{sum} = [draw,circle,inner sep=0mm,minimum size=2mm]
\tikzstyle{connector} = [->,thick]
\tikzstyle{line} = [thick]

\tikzstyle{branch} = [circle,inner sep=0pt,minimum size=1mm,fill=black,draw=black]

\tikzstyle{helperbranch} = [rectangle,inner sep=0pt,minimum size=0.1mm,fill=black,draw=black]

\tikzstyle{guide} = []

\tikzstyle{background}=[rectangle,
fill=blue!10,
inner sep=0.3cm,
rounded corners=5mm]

\tikzstyle{backgroundtwo}=[rectangle,
fill=red!10,
inner sep=0.4cm,
rounded corners=5mm]

\makeatletter

\resizebox{1\columnwidth}{!}{
\begin{tikzpicture}[>=latex, scale=0.7, auto, >=stealth']	

\small

\matrix[ampersand replacement=\&, row sep=0.2cm, column sep=0.5cm] {
	\& \&  \node[blockfixwidth] (opt1) {\begin{tabular}{c}
			Slack Variables \\ Computation \eqref{eq:pcbf}
		\end{tabular}};     \&  \\
	\node[block] (sys) {System \eqref{eq:sys}}; \&  \node[helperbranch] (hbx) {}; \& \node[blockfixwidth] (perf) {Performance Controller};  \&  \&  \\
	\&  \&   \& \node[helperbranch] (hb1){}; \\
	  \& \&  \node[block] (opt2) { Soft Constrained Predictive Safety Filter \eqref{eq:pcbf_sf}};\& \\
};

\draw [line](sys)--(hbx);
\draw [connector] (hbx) -- node {$x(k)$} (perf);
\draw [connector,] (hbx) |-node[near start]{} (opt1.west);
\draw [line] (opt1.east) -| node[near start]{$\xi^*_{i|k}$} (hb1.north);
\draw [line] (perf.east) -|  node[near start]{$u_p(k)$}  ([xshift=-2mm]hb1.north);

\draw [connector] (hb1)++(-0.2,0) |- node {}  ([yshift=+2mm]opt2);
\draw [connector] (hb1) |- node {}  ([yshift=-2mm]opt2);
\draw [connector] (opt2.west) -| node [near end] {$u_s(k)$} (sys.south);
\draw [connector] (hbx.south) |-++(0,-24pt) -|(opt2.north);

\end{tikzpicture}
}
    \caption{Outline of the PCBF-based safety filter framework in \cite{wabersich22}. Given the current state, the optimal slack variables and the performance input are computed, e.g., by evaluating a learning-based policy. Subsequently, the obtained values are used in the safety filter problem \eqref{eq:pcbf_sf}.}
    \label{fig:pcbf_sf}
    \vspace{-0.8cm}
\end{figure}
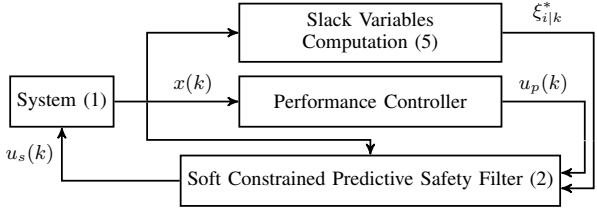

Finally, in order to derive the desired safety guarantees, the non-negative slack variables $\xi^*_{i|k}$ that are used in \eqref{eq:pcbf_sf} are not optimization variables, but are computed prior to \eqref{eq:pcbf_sf} in a second optimization problem minimizing the slack values:
\begin{subequations}
\label{eq:pcbf}
\begin{align}
 \hpb(x(k)) := \min_{u_{i|k}, \xi_{i|k}} \; \; & \alpha_f \xi_{N|k} + \sum_{i=0}^{N-1}\lvert \lvert{\xi_{i|k}}\rvert \rvert \label{eq:cbf_pcbf} \\
\text{s.t.}\; \; &\forall\, i = 0,\dots,N-1, \nonumber \\
& x_{0|k} = x(k), \label{eq:slack_init_pred} \\
& x_{i+1|k} = f(x_{i|k}, u_{i|k}), \\
& u_{i|k} \in \mathcal{U}, \label{eq:slack_input_constraint}\\
& x_{i|k} \in \overline{\mathcal{X}}_i(\xi_{i|k}), \; 0 	\leq \xi_{i|k}, \label{eq:pcbf_state_constraint}\\
& h_f(x_{N|k}) \leq \xi_{N|k}, \; 0 \leq \xi_{N|k},
\end{align}
\end{subequations}
using a terminal slack weight $\alpha_f>0$.
The resulting two step procedure of first computing the minimal slacks via \eqref{eq:pcbf}, followed by the computation of the input in \eqref{eq:pcbf_sf}, which is applied to the system, is shown schematically in Figure~\ref{fig:pcbf_sf}. It is shown in \cite{wabersich22}, that the optimal value function of \eqref{eq:pcbf}, denoted as $\hpb(x(k))$, fulfills the conditions of a CBF according to Definition~\ref{def:CBF} and is therefore referred to as a PCBF, as stated in the following Theorem. 

\begin{theorem}[Theorem III.6 in~\cite{wabersich22}]\label{thm:1}
Let $\mathcal{U}$ and $\overline{\mathcal{X}}_0(\xi)$ be compact for all $ 0 \leq \xi < \infty $ and $h_f$ be a control barrier function according to Definition~\ref{def:CBF}, with a corresponding safe set $\mathcal{S}_f \subset \overline{\mathcal{X}}_{N-1}(0)$, and continuous on $\mathbb{R}^n$. The minimum of \eqref{eq:pcbf} exists and if $\alpha_f<\infty$ is chosen sufficiently large, then the optimal value function $\hpb(x(k))$ defined in \eqref{eq:cbf_pcbf} is a control barrier function according to Definition~\ref{def:CBF} with domain $\Dpb := \{ x \in \mathbb{R}^n \mid \hpb(x) \leq \alpha_f \gamma_f \}$ and safe set $\Spb := \{ x \in \mathbb{R}^n \mid \hpb(x) = 0  \}$.
\end{theorem}

It follows from Theorem \ref{thm:1} that the inputs obtained from the framework are safe, i.e., the optimal solutions  $u_{0|k}^* \in \KCBF$. Additionally, in situations where predicted states are forced to leave the tightened state constraints $\overline{\mathcal{X}}_i$, e.g., due to large, unexpected disturbances or unsafe initial conditions, asymptotic stability of the slack variables $\xi_{i|k}^*$ with respect to the origin is ensured according to \cite[Theorem III.4]{wabersich22}. Therefore, all states $x\in \Dpb\setminus\Spb$ converge to the safe set $\Spb$, which is a subset of the implicitly defined feasible set of the hard constrained safety filter problem, i.e., \eqref{eq:pcbf_sf} with slack variables and additional tightenings $\Delta_i$ all set to zero,  which we denote as $\mathcal{X}_{\mathrm{feas}}$. 

\section{APPROXIMATE PCBF}\label{sec:approximate_sf}
In this section, we introduce the proposed method for approximating the PCBF-based safety filter algorithm. 
We first reformulate the predictive safety filter~\eqref{eq:pcbf_sf} in terms of a decrease condition on the predictive barrier function over one time step.
The resulting formulation is not directly implementable as an explicit formulation for the PCBF is generally not available. However, it provides the foundation for integrating an approximation $\hat{h}_{\mathrm{PB}}$, that is obtained by sampling the PCBF, into the approximate safety filter formulation by means of suitable choices for an implementable approximate decrease function.
We further show that any CBF $h$ on a compact domain $\mathcal{D}$, which fulfills the conditions in Definition~\ref{def:CBF} with a continuous decrease function $\Delta h(x)$ on $\mathcal{D}$, admits a class $\mathcal{K}$\footnote{A class $\mathcal{K}$ function $\gamma$ is continuous, strictly increasing and $\gamma(0) {=} 0$.} 
decrease function. Such a class $\mathcal{K}$ decrease enables a simplified convergence analysis compared to the positive continuous decrease $\Delta h$.
\subsection{One step predictive safety filter}
It is shown in \cite{wabersich22} that a continuous CBF decrease function $\Delta \hpb(x)$ according to \eqref{eq:rate_cond_cbf} exists for the optimal value function $\hpb(x)$ of the slack computation \eqref{eq:pcbf} under application of the safety filter policy, which implies convergence of the system states to $\Spb$ according to \cite[Theorem III.4]{wabersich22}. 
In order to make use of an approximation of the PCBF in Section~\ref{sec:approximate_pcbf_sf}, we first consider the following one step safety filter policy, which is in standard CBF form as considered, e.g., in \cite{ames2019} for the continuous-time case:
\begin{subequations}
\label{eq:true_cbf_sf}
\begin{align}
 \hspace{-0.2cm} \pi(x,u_p)\in&\textup{ arg}\min_{u \in \mathcal{U}} \;  \lvert \lvert{u_p - u}\rvert \rvert  \\
  &\textup{ s.t. }  \hpb(f(x,u)) - \hpb(x) \leq - \Delta \hpb(x).
  \label{eq:true_cbf_sf_constraint}
\end{align}
\end{subequations}

It follows from \cite[Theorem III.4]{wabersich22}, that this safety filter formulation gives convergence guarantees similar to solving \eqref{eq:pcbf} and then \eqref{eq:pcbf_sf} as at every time step, for any $x(k)\in\Dpb$, the safety filter policy $\pi(x(k),u_p(k))\in\mathcal{U}_{\textup{CBF}}(x(k))$. 
Compared to the two step formulation in \cite{wabersich22}, only a single, scalar constraint needs to be imposed in \eqref{eq:true_cbf_sf}.
While constraint~\eqref{eq:true_cbf_sf_constraint} is typically non-convex as $\hpb$ is generally a non-convex function, 
\eqref{eq:true_cbf_sf_constraint} is independent of the time horizon which significantly reduces the number of optimization variables, i.e., the optimization variables are $u\in\mathcal U\subset \mathbb{R}^m$ compared to $\{u_{i|k}\}\in\mathcal U^N\subset \mathbb{R}^{Nm}$. 

The challenge of using \eqref{eq:true_cbf_sf} is the requirement that both $\hpb$ and $\Delta \hpb$ are available in closed form, yet $\hpb$ is the optimal value function of an optimization problem for which a closed form solution generally is not available. However, the exact value of $\hpb$ can be sampled for any state $x\in \Dpb$ by solving \eqref{eq:pcbf}. This motivates the use of learning-based methods to approximate $\hpb$ and obtain an explicit formulation which can be used in \eqref{eq:true_cbf_sf}. However, as the decrease function $\Delta \hpb$ is not unique and cannot be easily sampled, a suitable substitute needs to be chosen.

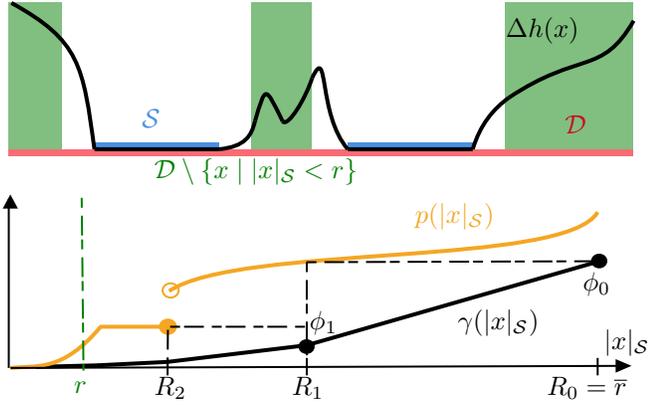
\begin{figure}
    \centering
    \vspace{0.2cm}
    \tikzset{every picture/.style={line width=0.75pt}} 

\begin{tikzpicture}[x=0.75pt,y=0.75pt,yscale=-0.8,xscale=0.9]

\draw [color={rgb, 255:red, 0; green, 0; blue, 0 }  ,draw opacity=1 ][line width=1.5]    (150.5,279.33) -- (115,280.33) ;
\draw  [draw opacity=0][fill={rgb, 255:red, 0; green, 128; blue, 0 }  ,fill opacity=0.5 ] (450,50) -- (378.13,50) -- (378.13,144) -- (450,144) -- cycle ;
\draw  [draw opacity=0][fill={rgb, 255:red, 0; green, 128; blue, 0 }  ,fill opacity=0.5 ] (270,50) -- (236,50) -- (236,144) -- (270,144) -- cycle ;
\draw  [draw opacity=0][fill={rgb, 255:red, 0; green, 128; blue, 0 }  ,fill opacity=0.5 ] (130,50) -- (100,50) -- (100,144) -- (130,144) -- cycle ;
\draw  [draw opacity=0][fill={rgb, 255:red, 74; green, 144; blue, 226 }  ,fill opacity=1 ] (148,138.67) -- (218,138.67) -- (218,143) -- (148,143) -- cycle ;
\draw  [draw opacity=0][fill={rgb, 255:red, 74; green, 144; blue, 226 }  ,fill opacity=1 ] (290,138.67) -- (360,138.67) -- (360,143) -- (290,143) -- cycle ;
\draw  [draw opacity=0][fill={rgb, 255:red, 249; green, 108; blue, 116 }  ,fill opacity=1 ] (100,143.1) -- (450,143.1) -- (450,147.22) -- (100,147.22) -- cycle ;
\draw [line width=1.5]    (218,143) .. controls (232.62,141.29) and (236,134.56) .. (238,125.22) ;
\draw [line width=1.5]    (148,143) -- (218,143) ;
\draw [line width=1.5]    (253.48,125.43) .. controls (243.48,102.43) and (244,102.56) .. (238,125.22) ;
\draw [line width=1.5]    (253.48,125.43) .. controls (256.48,129.43) and (266.48,110.71) .. (270,98.05) ;
\draw [line width=1.5]    (290,143) -- (360,143) ;
\draw [line width=1.5]    (270,98.05) .. controls (279.48,75.43) and (273,121.05) .. (290,143) ;
\draw [line width=1.5]    (101.48,50.71) .. controls (140,74.48) and (141.62,90.48) .. (148,143) ;
\draw [line width=1.5]    (360,143) .. controls (364.19,127.52) and (371.19,117.95) .. (380.19,110.52) ;
\draw [line width=1.5]    (380.19,110.52) .. controls (420.19,80.52) and (432.19,96.95) .. (450.19,61.95) ;
\draw [color={rgb, 255:red, 245; green, 166; blue, 35 }  ,draw opacity=1 ][line width=1.5]    (100.96,281) .. controls (115.58,279.29) and (130,283.84) .. (151.5,255) ;
\draw [color={rgb, 255:red, 245; green, 166; blue, 35 }  ,draw opacity=1 ][line width=1.5]    (150.87,255) -- (189.19,255) ;
\draw  [color={rgb, 255:red, 245; green, 166; blue, 35 }  ,draw opacity=1 ] [fill={rgb, 255:red, 245; green, 166; blue, 35 }  ,fill opacity=1 ](184.69,255) .. controls (184.69,252.51) and (186.71,250.5) .. (189.19,250.5) .. controls (191.68,250.5) and (193.69,252.51) .. (193.69,255) .. controls (193.69,257.48) and (191.68,259.5) .. (189.19,259.5) .. controls (186.71,259.5) and (184.69,257.48) .. (184.69,255) -- cycle ;
\draw  [color={rgb, 255:red, 245; green, 166; blue, 35 }  ,draw opacity=1 ] (186.19,232.1) .. controls (186.19,229.61) and (188.21,227.6) .. (190.69,227.6) .. controls (193.18,227.6) and (195.19,229.61) .. (195.19,232.1) .. controls (195.19,234.58) and (193.18,236.6) .. (190.69,236.6) .. controls (188.21,236.6) and (186.19,234.58) .. (186.19,232.1) -- cycle ;
\draw [color={rgb, 255:red, 245; green, 166; blue, 35 }  ,draw opacity=1 ][line width=1.5]    (189.87,232.53) .. controls (229.87,202.53) and (412,217.67) .. (430,182.67) ;
\draw [line width=0.75]    (100.96,281) -- (446,279.68) ;
\draw [shift={(449,279.67)}, rotate = 179.78] [fill={rgb, 255:red, 0; green, 0; blue, 0 }  ][line width=0.08]  [draw opacity=0] (8.93,-4.29) -- (0,0) -- (8.93,4.29) -- cycle    ;
\draw    (100,281) -- (100,174.53) ;
\draw [shift={(100.87,171.53)}, rotate = 89.95] [fill={rgb, 255:red, 0; green, 0; blue, 0 }  ][line width=0.08]  [draw opacity=0] (8.93,-4.29) -- (0,0) -- (8.93,4.29) -- cycle    ;
\draw [color={rgb, 255:red, 0; green, 128; blue, 0 }  ,draw opacity=1 ] [dash pattern={on 3.75pt off 3pt on 7.5pt off 1.5pt}]  (140.87,173.67) -- (141.87,282.67) ;
\draw  [dash pattern={on 3.75pt off 3pt on 7.5pt off 1.5pt}]  (189,280.67) -- (189.19,255) ;
\draw  [dash pattern={on 3.75pt off 3pt on 7.5pt off 1.5pt}]  (267,214.27) -- (267,279.67) ;
\draw    (430,274.67) -- (430,285.67) ;
\draw    (267,274.67) -- (267,285.67) ;
\draw    (189,274.67) -- (189,285.67) ;
\draw [color={rgb, 255:red, 0; green, 0; blue, 0 }  ,draw opacity=1 ][line width=1.5]    (431,213.67) -- (267,266.67) ;
\draw  [dash pattern={on 3.75pt off 3pt on 7.5pt off 1.5pt}]  (266.87,214.27) -- (431,213.67) ;
\draw  [dash pattern={on 3.75pt off 3pt on 7.5pt off 1.5pt}]  (266,255) -- (189.19,255) ;
\draw  [color={rgb, 255:red, 0; green, 0; blue, 0 }  ,draw opacity=1 ][fill={rgb, 255:red, 0; green, 0; blue, 0 }  ,fill opacity=1 ] (262.87,267.53) .. controls (262.87,265.32) and (264.66,263.53) .. (266.87,263.53) .. controls (269.08,263.53) and (270.87,265.32) .. (270.87,267.53) .. controls (270.87,269.74) and (269.08,271.53) .. (266.87,271.53) .. controls (264.66,271.53) and (262.87,269.74) .. (262.87,267.53) -- cycle ;
\draw [color={rgb, 255:red, 0; green, 0; blue, 0 }  ,draw opacity=1 ][line width=1.5]    (267,266.67) -- (189,277.19) ;
\draw [color={rgb, 255:red, 0; green, 0; blue, 0 }  ,draw opacity=1 ][line width=1.5]    (189,277.19) -- (150,279.33) ;
\draw  [color={rgb, 255:red, 0; green, 0; blue, 0 }  ,draw opacity=1 ][fill={rgb, 255:red, 0; green, 0; blue, 0 }  ,fill opacity=1 ] (426.87,213.53) .. controls (426.87,211.32) and (428.66,209.53) .. (430.87,209.53) .. controls (433.08,209.53) and (434.87,211.32) .. (434.87,213.53) .. controls (434.87,215.74) and (433.08,217.53) .. (430.87,217.53) .. controls (428.66,217.53) and (426.87,215.74) .. (426.87,213.53) -- cycle ;

\draw (432,255) node [anchor=north west][inner sep=0.75pt]   [align=left] {$\displaystyle |x|_{\mathcal{S}}$};
\draw (173,116) node [anchor=north west][inner sep=0.75pt]   [align=left] {$\displaystyle \textcolor[rgb]{0.29,0.56,0.89}{\mathcal{S}}$};
\draw (410,120) node [anchor=north west][inner sep=0.75pt]  [color={rgb, 255:red, 249; green, 108; blue, 116 }  ,opacity=1 ] [align=left] {$\displaystyle \textcolor[rgb]{0.82,0.01,0.11}{\mathcal{D}}$};
\draw (180,147) node [anchor=north west][inner sep=0.75pt]  [color={rgb, 255:red, 0; green, 100; blue, 0 }  ,opacity=1 ] [align=left] {$\displaystyle \textcolor[rgb]{0,0.5,0}{\mathcal{D}\setminus\{x\mid|x|_{\mathcal{S}} < r\}}$};
\draw (377,58) node [anchor=north west][inner sep=0.75pt]    {$\Delta h( x)$};
\draw (326,175) node [anchor=north west][inner sep=0.75pt]    {$\textcolor[rgb]{0.96,0.65,0.14}{p( |x|_{\mathcal{S}})}$};
\draw (400,285) node [anchor=north west][inner sep=0.75pt]    {$R_{0}=\overline{r}$};
\draw (350,242) node [anchor=north west][inner sep=0.75pt]  [color={rgb, 255:red, 0; green, 0; blue, 0 }  ,opacity=1 ]  {$\gamma ( |x|_{\mathcal{S}})$};
\draw (267,243) node [anchor=north west][inner sep=0.75pt]    {$\phi_{1}$};
\draw (420,219) node [anchor=north west][inner sep=0.75pt]    {$\phi_{0}$};
\draw (257,285) node [anchor=north west][inner sep=0.75pt]    {$R_{1}$};
\draw (180,285) node [anchor=north west][inner sep=0.75pt]    {$R_{2}$};
\draw (135,288) node [anchor=north west][inner sep=0.75pt]  [color={rgb, 255:red, 144; green, 225; blue, 58 }  ,opacity=1 ]  {$\textcolor[rgb]{0,0.5,0}{r}$};

\end{tikzpicture}
    \vspace{-0.5cm}
    \caption{Illustration of the proof of Proposition~\ref{prob:delta_h_lower_bound_kfunc}. The continuous function $\Delta h$ is $0$ on the compact set $\mathcal{S}$ and strictly positive on $\mathcal{D}\setminus\mathcal{S}$. A non-decreasing, potentially non-continuous, lower bound $p$ is constructed, which is in turn lower bounded by a piecewise linear function connecting the sequence of points $(R_k,\phi_k)$.}
    \label{fig:classK}
    \vspace{-0.5cm}
\end{figure}

In the following, we show that for any CBF $h(x)$ which fulfills Definition~\ref{def:CBF}, a class $\mathcal{K}$ lower bound exists for any continuous decrease function $\Delta h(x)$ on the compact set $\mathcal{D}$, which enables a simplified convergence analysis. The lower bound construction is illustrated in Figure~\ref{fig:classK}.

\begin{proposition}
\label{prob:delta_h_lower_bound_kfunc}
Consider a CBF $h(x)$ according to Definition~\ref{def:CBF} and a corresponding decrease function $\Delta h(x)$ which is continuous and strictly positive for all $x\in\mathcal{D}\setminus\mathcal{S}$ and $\Delta h(x)=0$ for all $x\in\mathcal{S}$. Then there exists a class $\mathcal{K}$ function $\gamma: [0,\overline{r}] \to \mathbb{R}_{\geq 0}$, with $\overline{r}:= \max_{x\in \mathcal{D}}|x|_{\mathcal{S}}$ and $|x|_{\mathcal{S}} := \inf\limits_{y\in \mathcal{S}} || x - y|| $, such that $\Delta h(x) \geq \gamma(|x|_{\mathcal{S}})$ for all $x \in \mathcal{D}$.
\end{proposition}
\begin{proof}
Define the auxiliary function $p\!:\! [0,\overline{r}] {\to} \mathbb{R}_{\geq0}$ as
\begin{align}
\label{eq:prop_1_min_bound}
    \hspace{-0.1cm}p(r) \!:=\! 
    \begin{cases}
    0 &\hspace{-0.5cm} \text{if } r = 0, \\
    & \\
    \parbox{4.7cm}{\footnotesize $\min\limits_y \;\; \Delta h(y) \\
    \text{s.t.} \;\;  y \in \mathcal{D}\setminus \{y \in \mathbb{R}^n \mid |y|_{\mathcal{S}} < r\}$}& \hspace{-0.5cm}\mathrm{otherwise}.
    \end{cases}
\end{align}
Since $\Delta h$ is a continuous function and $\mathcal{D} \setminus \{y \in \mathbb{R}^n \mid |y|_{\mathcal{S}} < r\}$ is a compact set, the above minimum is attained for all $r\in(0,\overline{r}]$. 
It holds that
\begin{align*}
p(|x|_{\mathcal{S}}) \leq  \Delta h(x), \; \forall x \in \mathcal{D},
\end{align*}
which follows directly from the above minimization as $x$ is a feasible point if $|x|_{\mathcal{S}} > 0$ and the fact that $\Delta h = 0, \; \forall x \in \mathcal{S}$. Additionally, it holds that $p$ is non-decreasing with increasing $r$, since $\bar r > r$ implies $\mathcal{D}\setminus\{y \in \mathbb{R}^n \mid |y|_{\mathcal{S}} < \bar r\}\subseteq\mathcal{D}\setminus\{y \in \mathbb{R}^n \mid |y|_{\mathcal{S}} < r\}$, i.e., the domain in the constrained minimization problem (second case in \eqref{eq:prop_1_min_bound}) is non-increasing and $\Delta h(x) = 0,  \; \forall x \in \mathcal{S}$.
As $p(r)$ is not necessarily continuous for non-convex sets $\mathcal S$, we construct a class $\mathcal{K}$ function $\gamma : [0,\overline{r}] \to \mathbb{R}_{\geq0}$ which is a lower bound of $p$ on $\mathcal{D}$ similarly to the result in~\cite[Lemma 2.5]{clarke98}.
Consider the partitions $(R_{k+1},R_k]$ on the interval $(0,\overline{r}]$ with boundary points $R_k := \overline{r}2^{-k} , \mathrm{for} \; k = 0, 1, 2, \dots$ and define $\phi_k:=2^{-k} p(R_{k+1})$.
Note that $\phi_{k+1} < \phi_{k}$, since $p(r)$ is non-decreasing and $2^{-k-1} < 2^{-k}$ $\forall k \geq 0$. In addition, it holds that $\phi_k\leq p(r), \forall r \in (R_{k+1},R_k]$ as $\phi_k=2^{-k} p(R_{k+1}) \leq p(R_{k+1})$.

We construct the function
\begin{align*}
\gamma(r) = \begin{cases}
    0 & \text{if } r=0, \\
    \phi_{k+1} + \frac{\phi_{k} - \phi_{k+1}}{(R_{k}-R_{k+1})}(r-R_{k+1}),& r\in(R_{k+1},R_k],
\end{cases}
\end{align*}
which is piecewise affine on successive intervals $r \in (R_{k+1}, R_{k}]$, continuous, as $\lim_{k\rightarrow \infty} \phi_k = 0$, strictly increasing and is therefore a class $\mathcal{K}$ function. Furthermore, it holds that $\gamma(r) \leq p(r)$, which implies $\gamma(|x|_{\mathcal{S}}) \leq \Delta h(x) \; \forall x\in\mathcal{D}$ as desired.
\end{proof}

The provided class $\mathcal{K}$ lower bound on the decrease implies that under the application of $\pi(x(k),u_p(k))$, at every time step and for all $x(k)\in\Dpb$, the PCBF decrease is given by $\hpb(f(x(k),u(k)))-\hpb(x(k))\leq -\Delta\hpb(x(k))\leq -\gamma(|x(k)|_{\Spb})$, which implies convergence to $\Spb$.
Furthermore, this result enables showing convergence to a neighborhood of the safe set defined by the PCBF $\hpb$ of the approximate safety filter scheme proposed in Section~\ref{sec:approximate_pcbf_sf}. 

\begin{remark}
Note that the result from Proposition \ref{prob:delta_h_lower_bound_kfunc} can also be used to simplify the convergence proof in~\cite[Theorem III.4]{wabersich22}.
\end{remark}



\subsection{Learning the CBF} \label{subsec:data_gen}
An approximation of $\hpb$, denoted by $\approxhpb$, is obtained by training a regression model on a data-set $\mathbb{D}=\{(x_i,\hpb(x_i)\}_{i=1}^{n_{\mathrm{data}}}$, where $n_{\mathrm{data}}$ is the number of sampled points. The sampled state space $\Omega$ is chosen to cover scenarios which are task specific and deemed physically realistic. One possible choice for the sample space is $\Omega = \{ x \in \mathbb{R}^n \mid \hpb(x) \leq H\}$ with $0<H\leq \alpha_f\gamma_f$, where a rejection sampler similar to \cite{chen22} can be used to generate~$\mathbb{D}$. Note that $\hpb$ is always decreasing along individual trajectories in $\Omega\subseteq\Dpb$ under the PCBF-based safety filter algorithm, which implies that the specified $\Omega$ is an invariant set for inputs $u\in \KCBF$. 
In this work, we focus on neural networks $\approxhpb(x) = \mathrm{NN}(x;\theta)$ whose weights are denoted as $\theta$. In principle, the proposed algorithm can be used with other regression techniques as long as they allow to be integrated in an optimization-based framework. We assume that $\approxhpb$ fulfills the following assumption.

\begin{assumption}
\label{ass:uniform_error}
The approximation error is uniformly bounded for the trained regressor, i.e., it holds that $|\hpb(x)-\approxhpb(x)| \leq \epsilon_h$ for all $x \in \Dpb$. In other words, it holds that $\approxhpb(x) =\hpb(x) + e_{h}(x) $, where the error $e_{h}(x)$ is bounded by $|e_h(x)|\leq \epsilon_h$.
\end{assumption}

Since $\hpb$ is continuous and defined on a compact set $\Dpb \subset \mathbb{R}^n$ (see, \cite[Theorem III.6]{wabersich22}), the universal approximation theorem in~\cite{hornik1989} guarantees the existence of a neural network, which can approximate $\hpb$ with an arbitrary small approximation error. Note that if the chosen regression results in a continuous function $\approxhpb$ on $\Dpb$, then a bounded approximation error always exists on the compact domain $\Dpb$.
Furthermore, Assumption~\ref{ass:uniform_error} can be validated, by extracting a statistical estimate obtained from a certain number of sampled points, as done, e.g., in~\cite{hertneck2018}.  

\subsection{Safety filter using PCBF approximation}\label{sec:approximate_pcbf_sf}
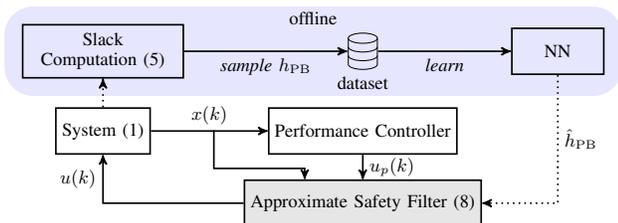
\begin{figure}[h]
    \centering
    \pgfdeclarelayer{background}
\pgfdeclarelayer{foreground}
\pgfsetlayers{background, main, foreground}

\tikzstyle{block} = [draw,rectangle,thick,minimum height=2.5em,minimum width=5em]
\tikzstyle{blockfilled} = [draw,rectangle, fill=black!10, thick,minimum height=2.5em,minimum width=5em]

\tikzstyle{sum} = [draw,circle,inner sep=0mm,minimum size=2mm]
\tikzstyle{connector} = [->,thick]
\tikzstyle{line} = [thick]

\tikzstyle{branch} = [circle,inner sep=0pt,minimum size=1mm,fill=black,draw=black]

\tikzstyle{helperbranch} = [rectangle,inner sep=0pt,minimum size=0.1mm,fill=black,draw=black]

\tikzstyle{guide} = []

\tikzstyle{bluebackground}=[rectangle,
fill=blue!10,
inner sep=0.3cm,
rounded corners=5mm]

\tikzstyle{backgroundtwo}=[rectangle,
fill=red!10,
inner sep=0.4cm,
rounded corners=5mm]

\makeatletter
\tikzset{
	database/.style={
		path picture={
			\draw (0, 1.5*\database@segmentheight) circle [x radius=\database@radius,y radius=\database@aspectratio*\database@radius];
			\draw (-\database@radius, 0.5*\database@segmentheight) arc [start angle=180,end angle=360,x radius=\database@radius, y radius=\database@aspectratio*\database@radius];
			\draw (-\database@radius,-0.5*\database@segmentheight) arc [start angle=180,end angle=360,x radius=\database@radius, y radius=\database@aspectratio*\database@radius];
			\draw (-\database@radius,1.5*\database@segmentheight) -- ++(0,-3*\database@segmentheight) arc [start angle=180,end angle=360,x radius=\database@radius, y radius=\database@aspectratio*\database@radius] -- ++(0,3*\database@segmentheight);
		},
		minimum width=2*\database@radius + \pgflinewidth,
		minimum height=3*\database@segmentheight + 2*\database@aspectratio*\database@radius + \pgflinewidth,
	},
	database segment height/.store in=\database@segmentheight,
	database radius/.store in=\database@radius,
	database aspect ratio/.store in=\database@aspectratio,
	database segment height=0.15cm,
	database radius=0.25cm,
	database aspect ratio=0.35,
}

\resizebox{1\columnwidth}{!}{
\begin{tikzpicture}[scale=1, auto, >=stealth']	

	\small

	\matrix[ampersand replacement=\&, row sep=0.2cm, column sep=0.5cm] {
		 \node[block] (opt1) {\begin{tabular}{c}
		 		Slack \\ Computation \eqref{eq:pcbf}
		 \end{tabular}};  \&  \& \node[database, label=below:dataset] (data) {};  \& \node[block] (net) {NN};  \\
		\node[block] (sys) {System \eqref{eq:sys}}; \& \node[helperbranch] (hbx) {};  \& \node[block] (perf) {Performance Controller};  \&  \&  \\
		\&  \&   \& \node[helperbranch] (hb1){}; \\
		  \& \&  \node[blockfilled] (opt2) {Approximate Safety Filter \eqref{eq:nn_cbf_sf}};\& \\
	};

	\draw [connector] (sys) -- node {$x(k)$} (perf);
	\draw [connector, dotted] (sys) |- (opt1.south);
	\draw [connector] (perf.south) -- node {$u_p(k)$} (opt2.north);
	\draw [connector] (opt1.east) -- node[below] {\textit{sample} $h_{\mathrm{PB}}$}(data.west);
	\draw [connector] (opt2.west) -| node [near end] {$u(k)$} (sys.south);
	\draw [connector] (data.east) -- node[below] {\textit{learn}}(net.west);
	\draw [connector, dotted] (net.south) |- node[right, near start] {$\hat{h}_{\mathrm{PB}}$}(opt2.east);
	\draw [connector] (hbx.south) |-++(0,-18pt) -|(  [xshift=-1cm] opt2.north);


%
	
	\begin{pgfonlayer}{background}
	\node [ bluebackground, label={[shift={(0,-3ex)}]above:offline},
	fit= (opt1) (data) (net)] {};
	\end{pgfonlayer}
	
	\end{tikzpicture}
}
    \vspace{-0.9cm} 
    \caption{Illustration of the proposed approximation of the PCBF. 
    The optimization problem \eqref{eq:pcbf}, which computes the optimal slacks is only solved offline to sample data points of $\hpb$. Based on these samples, a neural network is trained so that an explicit formulation of the approximation $\ahpb$ of $\hpb$, can be used online in an approximate safety filter scheme.}
    \label{fig:approximate_pcbf_block}
\end{figure}

The learned neural network approximation $\approxhpb(x) = \mathrm{NN}(x;\hat{\theta})$ can be used in an approximate safety filter scheme by inserting the approximation into \eqref{eq:true_cbf_sf}, resulting in the approximate safety filter control law
\begin{subequations}
\label{eq:nn_cbf_sf}
\begin{align}
\hspace{-0.2cm}\hat{\pi}(x, u_p)\in & \textup{ arg} \min_{u \in \mathcal{U}} \; \; \lvert \lvert{u_p - u}\rvert \rvert  \\
& \textup{ s.t. } \approxhpb(f(x,u))- \approxhpb(x) \leq - \Delta\hat{h}_\mathrm{PB}(x),
\label{eq:nn_cbf_sf_state_constraints}
\end{align}
\end{subequations}
where $\Delta\approxhpb$ is a decrease function for the approximate CBF. The resulting safety filter scheme is illustrated in Figure~\ref{fig:approximate_pcbf_block}. In order to provide a safety filter formulation which guarantees feasibility, we consider the maximal approximate decrease, which is given by
\begin{equation}
\label{eq:approx_cbf_max_decrease}
    \Delta \approxhpb(x(k)) := \max_{u \in \mathcal{U}} \;\;   \approxhpb(x(k)) - \approxhpb(f(x(k),u)).
\end{equation}

Note that similar to the methods presented in \cite{wabersich22}, this choice of decrease function requires the solution to two optimization problems, which are however still independent of the prediction horizon. 
In Section~\ref{sec:numerical}, we discuss how a decrease function can be designed, which requires only one optimization problem, however at the expense of recursive feasibility guarantees.
\begin{theorem}\label{thm:max_decrease}
    Let $\approxhpb(x)$ be an approximate CBF which fulfills Assumption 1. Then it holds that the approximate safety filter policy \eqref{eq:nn_cbf_sf}, where $\Delta\approxhpb(x)$ is given by the solution of \eqref{eq:approx_cbf_max_decrease}, is feasible for $x\in\Dpb$ and for a sufficiently small error bound $\epsilon_h\geq0$, there exists an $R\geq0$ such that the system \eqref{eq:sys} under the application of $\hat{\pi}(x, u_p)$ converges to $\mathcal{C}:=\{x\in\mathbb{R}^n\mid |x|_{\Spb}\leq R\}$. Furthermore, if $\epsilon_h=0$, the closed-loop system converges to $\Spb$.
\end{theorem}
\begin{proof}
Feasibility of using \eqref{eq:approx_cbf_max_decrease} in \eqref{eq:nn_cbf_sf} follows from $u^* \in \mathrm{arg}\max_{u\in\mathcal{U}} \approxhpb(x)-\approxhpb(f(x,u))$ being a feasible solution in \eqref{eq:nn_cbf_sf}. Under Assumption~\ref{ass:uniform_error}, we have that for all $x\in\Dpb$, the maximum decrease \eqref{eq:approx_cbf_max_decrease} is bounded as follows: $\max_{u\in\mathcal{U}}\approxhpb(x)-\approxhpb(f(x,u))=\max_{u\in\mathcal{U}}\hpb(x)-\hpb(f(x,u))+e_h(x)-e_h(f(x,u))\geq \max_{u\in\mathcal{U}}\hpb(x)-\hpb(f(x,u)) - 2\epsilon_h\geq \Delta\hpb(x)-2\epsilon_h$, where the last inequality follows from the existence of $\Delta \hpb$ from Definition~\ref{def:CBF}. 
It holds that $\hpb(f(x,\hat{\pi}(x, u_p)))-\hpb(x) = \approxhpb(f(x,\hat{\pi}(x, u_p)))-\approxhpb(x)+e_h(f(x,\hat{\pi}(x, u_p)))-e_h(x)\leq -\Delta\approxhpb(x)+2\epsilon_h$, which, using Proposition~\ref{prob:delta_h_lower_bound_kfunc} and the previous maximum decrease bound, results in
\begin{align*}
  \hpb(f(x,\hat{\pi}(x, u_p))) - \hpb(x) & \leq - \gamma(|x|_{\Spb}) + 4 \epsilon_h, \label{eq:thm_max_decrease_final_rate} 
\end{align*}
which implies convergence to $\mathcal{C}$ for small enough $\epsilon_h$. Furthermore, it follows directly that for $\epsilon_h=0$, the closed-loop system converges to $\Spb$.
\end{proof}

While the approximate decrease in \eqref{eq:approx_cbf_max_decrease} benefits from strong theoretical guarantees, it substantially reduces the available input space for finding a safe input lying as close as possible to $u_p(k)$, which potentially renders~\eqref{eq:nn_cbf_sf} conservative. 
This conservativeness is especially problematic in the presence of approximation errors, where $\approxhpb(x)\neq 0$ for some $x \in \Spb$, resulting in unnecessary safety filter interventions. To counteract this effect, the decrease constraint \eqref{eq:nn_cbf_sf_state_constraints} can be relaxed by subtracting a small fixed tolerance parameter $\delta_{\mathrm{tol}} > 0 $, i.e., $\Delta \overline{h}_\mathrm{PB}(x):=\Delta\hat{h}_\mathrm{PB}(x)  - \delta_{\mathrm{tol}}$, resulting in the following CBF decrease at every time step:
\begin{align*}
  \hpb(f(x,\hat{\pi}(x, u_p))) - \hpb(x) & \leq - \gamma(|x|_{\Spb}) {+} 4 \epsilon_h{+}\delta_{\mathrm{tol}}.
\end{align*}
Such a relaxation preserves recursive feasibility and convergence to a neighborhood of $\Spb$, i.e., to a set $\overline{\mathcal{C}}\supset \mathcal{C}$, however for $\epsilon_h=0$ the system is no longer guaranteed to converge to $\Spb$. 
\begin{remark}
The choice of the tightening terms $\Delta_{i}$ in the constraints $\overline{\mathcal{X}}_i(\xi)$ directly affects the size of $\Spb$, since~$\Spb \subset \mathcal{X}_{\mathrm{feas}}$. Consequently, choosing $\Delta_i$ large enough and for small approximation errors, it holds that~$\mathcal{C}\subset \mathcal{X}_\mathrm{feas}$.
Additionally, instead of its use as an approximate CBF decrease, the optimization problem \eqref{eq:approx_cbf_max_decrease} can also be employed directly as a recovery strategy for system \eqref{eq:sys} whenever the unrelaxed and untightened problem \eqref{eq:pcbf_sf} becomes infeasible.
\end{remark}

\section{NUMERICAL EXAMPLE}\label{sec:numerical}
In this section, we demonstrate the proposed method for a simple autonomous driving application
, where we highlight its closed-loop behaviour and compare the solve time of our approach to the original PCBF-based algorithm in \cite{wabersich22}. The numerical example is implemented using CasADi~\cite{andersson2019} and IPOPT~\cite{biegler2009}. The example is run on a machine equipped with an AMD Ryzen 7 5800X CPU with 8 cores.

\paragraph*{System} We consider the same kinematic car model as used in~\cite{wabersich22}, with the dynamics given by
\begin{equation}\label{eq:example_sys_nonlinear}
\begin{alignedat}{2}
    \dot{y}_{\mathrm{off}} \; =  \; & (v+v_s) \sin (\Psi), &\quad \dot{\delta}\; =\; & u_1, \\
    \dot{\Psi}\; = \; & \frac{v + v_s}{L} \tan(\delta), &\quad \dot{v}\; =\; & u_2,
\end{alignedat}
\end{equation}
where $y_{\mathrm{off}}$ and $\Psi$ are the lateral position of the vehicle and the relative angle with respect to the center line, respectively, $\delta$ is the steering angle, and $v$ the velocity of the vehicle defined relative to a target velocity $v_s = 5 \; [m/s]$. The wheelbase is set to $L=5 \; [m]$. The system can be controlled using the steering input $u_1$ and the acceleration $u_2$. The state constraints on the system are given by $|y_{\mathrm{off}}| \leq 2, \; |\Psi|\leq \pi/4, \; |\delta| \leq \pi / 9$ and $-5 \leq v \leq 4$. The input constraints are set to $|u_1| \leq 1.4$, $ -5 \leq u_2 \leq  2$ and system~\eqref{eq:example_sys_nonlinear} is discretized using Euler forward with sampling time $T_s = 0.05$.

In the following, we demonstrate the capability of the proposed approximate safety filter to recover from large constraint violations in $y_{\mathrm{off}}$. We use the linear
controller $u_p(k) = K_p x(k)$, where $K_p \in \mathbb{R}^{2\times 4}$ is a matrix with all entries set to 10, i.e., the input is destabilizing and therefore not safe to apply. We use the design parameters $\mu_x = \mu_u = 0.1, \; \Delta_i = i \cdot 0.004$, $\alpha_f = 1000$, prediction horizon~$N=50$, and terminal control barrier function~$h_f(x) = x^T P x$.\footnote{The approach in~\cite[Section IV]{wabersich22} was used to calculate~$P$ with trace instead of logdet, resulting in~$\mathcal{D}_f {=} \{ x {\in} \mathbb{R}^n \mid \hpb(x) {\leq} \gamma_f \}$ with $\gamma_f {=} 0.01$.}

\begin{figure}[ht]
    \centering
    \vspace{-0.3cm}
    \includegraphics[width=\linewidth]{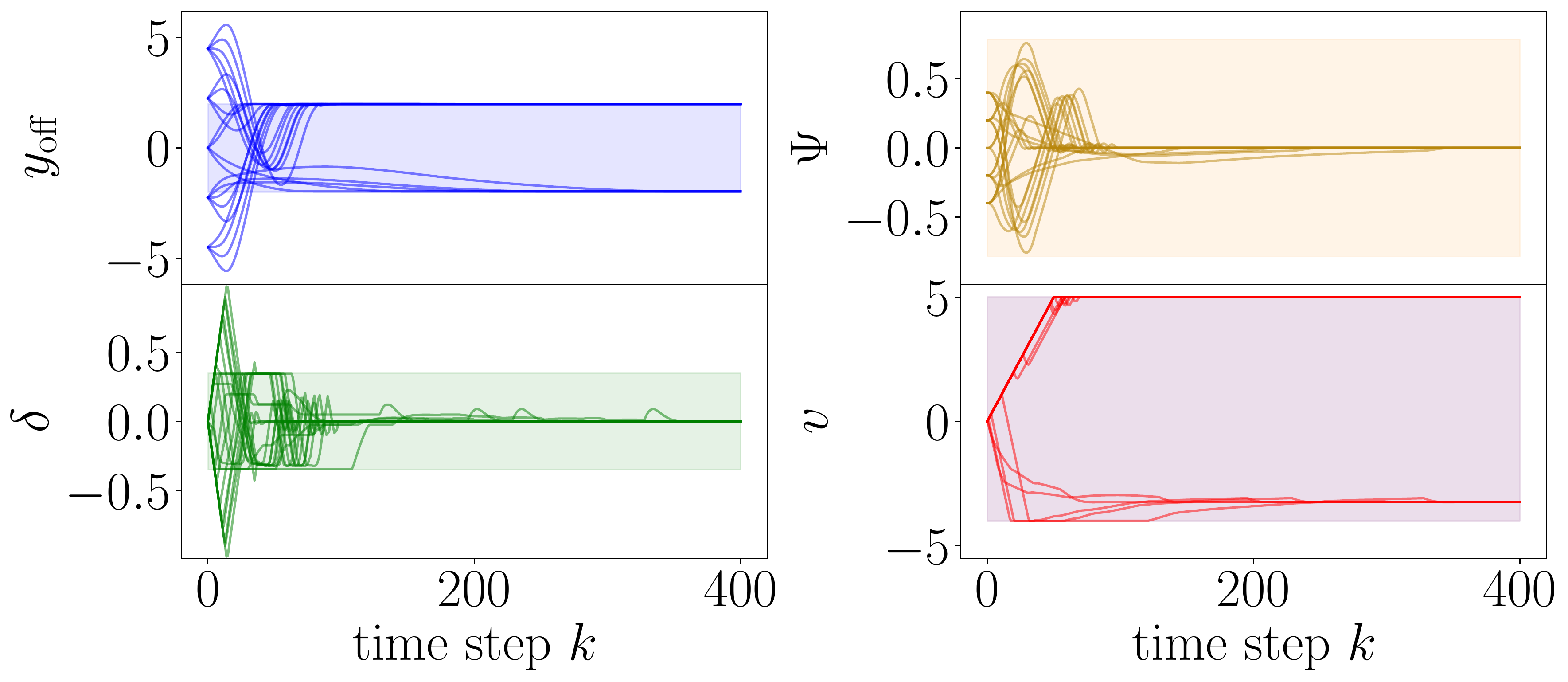}
    \vspace{-0.7cm}
    \caption{State trajectories for the closed-loop simulation of system \eqref{eq:example_sys_nonlinear} for multiple initial conditions using the PCBF-based safety filter in \cite{wabersich22}. The shaded background represents the state constraints $\mathcal{X}$.}
    \label{fig:nonlinear_pcbf_sf}
    \vspace{-0.2cm}
\end{figure}

\paragraph*{Learning}
We sample from the state space using the approach outlined in Section~\ref{subsec:data_gen}, where we set $H=100$ in conjunction with the rejection sampling scheme presented in~\cite{chen22}. In total, we generate 8.9 million data points using the rejection sampler and sample additional 2 million data points from a region corresponding to a scaled (by factor of 1.2) version of the state constraints to increase the diversity of the data set. The neural network is trained using a Huber loss to account for the fact that $\hpb$ can attain values over a wide range of magnitudes and we are generally more interested in a precise approximation close to the safe set, i.e., regions with small $\hpb$ values.
We use a neural network with three hidden layers, each containing 64 neurons, with a training time of around 4 hours. In order to analyze how the number of parameters in the neural network affects the solve time of our method, we additionally train a shallower network with two hidden layers. The \textit{softplus} activation function is used as the activation layer for both networks, which forces $\approxhpb$ to be positive.

\begin{figure}[t]
    \centering
    \vspace{0.15cm}
    \includegraphics[width=\linewidth]{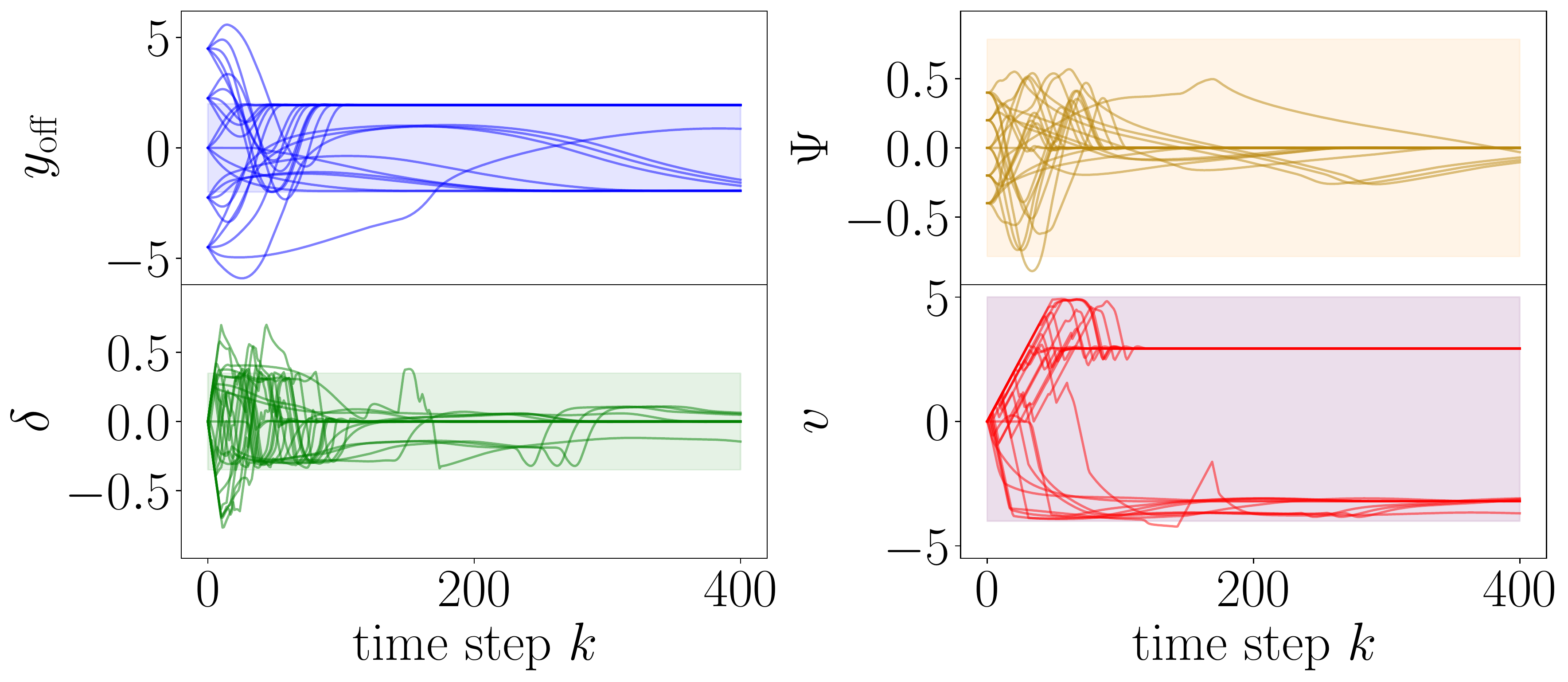}
    \vspace{-0.7cm}
    \caption{State trajectories for the closed-loop simulation of system \eqref{eq:example_sys_nonlinear} for multiple initial conditions using the CBF decrease function \eqref{eq:approx_cbf_max_decrease}. The CBF is approximated with a NN with 3 hidden layers and 64 neurons per layer.}
    \label{fig:nonlinear_maxdec_multi_ic}
    \vspace{-0.3cm}
\end{figure}

\begin{figure}[t]
    \centering
    \includegraphics[width=\linewidth]{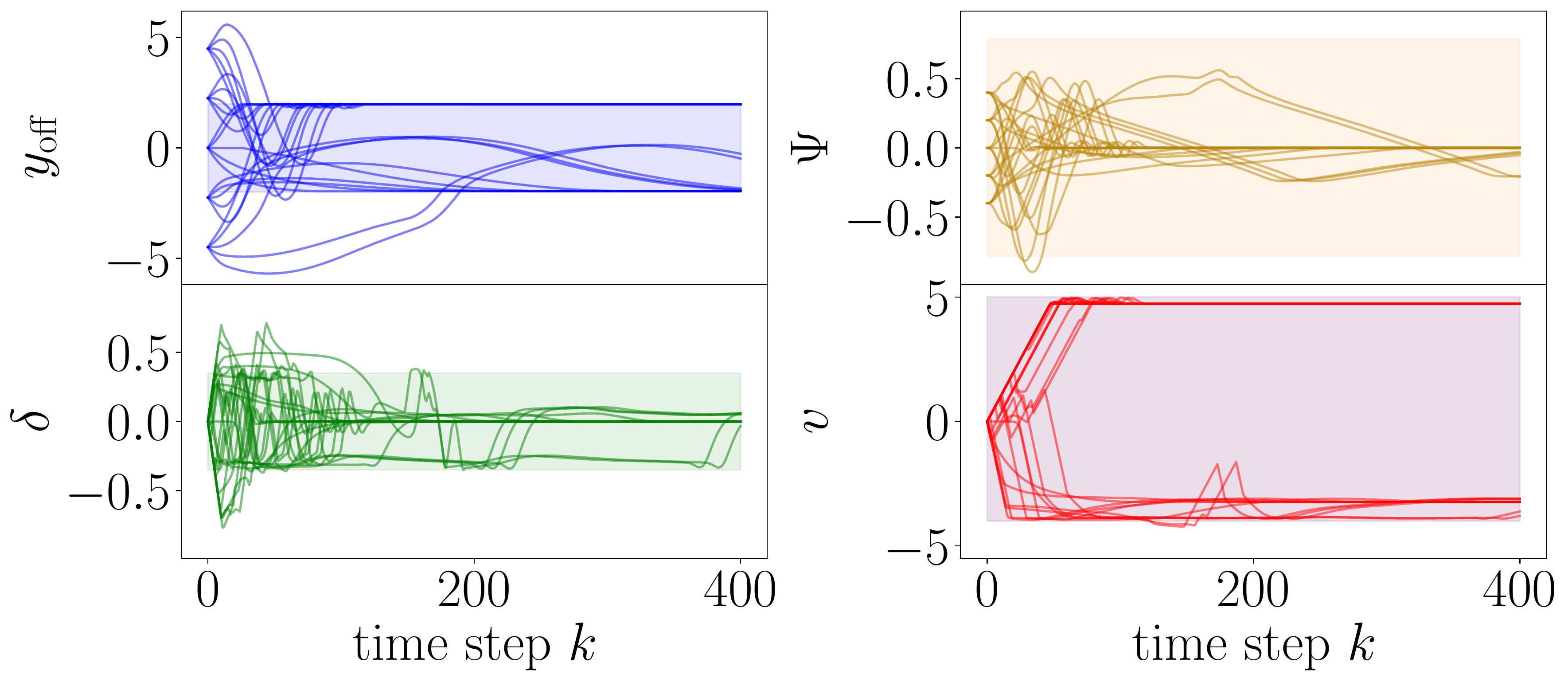}
    \vspace{-0.7cm}
    \caption{State trajectories for the closed-loop simulation of system \eqref{eq:example_sys_nonlinear} for multiple initial conditions using a hand-tuned class $\mathcal{K}$ function $\Delta\hat{h}_\mathrm{PB}(x) = 0.5 \approxhpb(x)$. The CBF is approximated with a NN with 3 hidden layers and 64 neurons per layer.}
    \label{fig:nonlinear_kinf_multi_ic}
    \vspace{-0.7cm}
\end{figure}

\begin{figure}[t]
    \centering
    \vspace{0.2cm}
    \includegraphics[width=\linewidth]{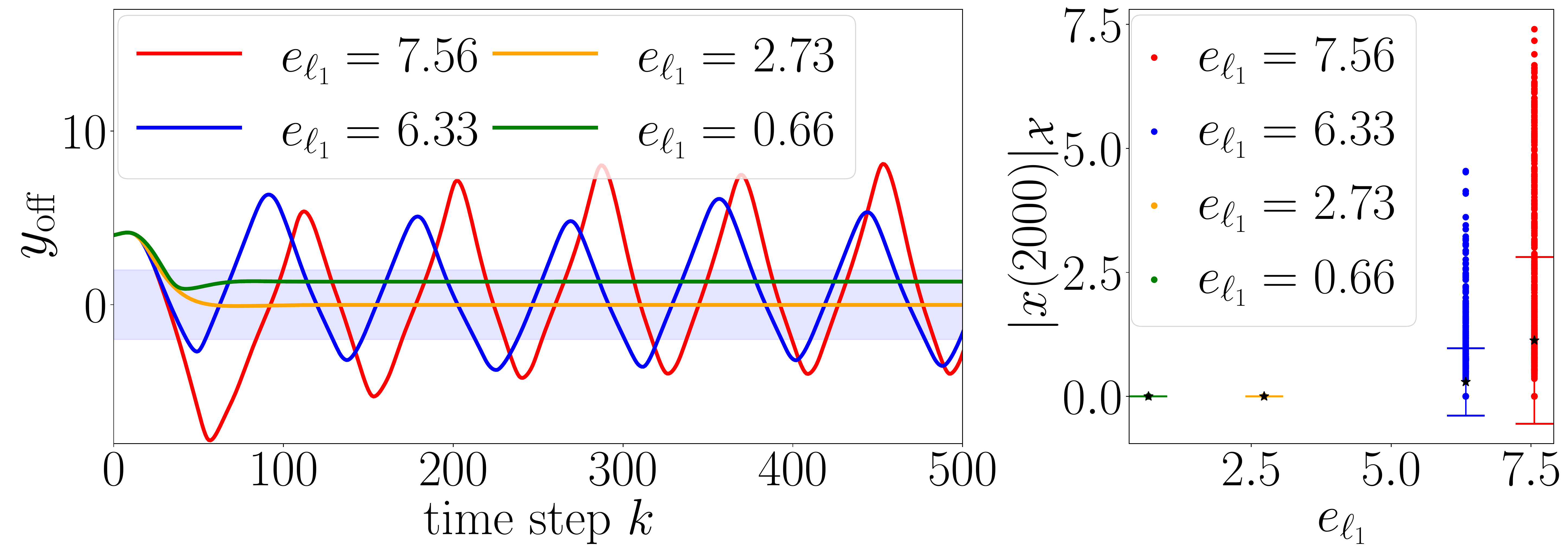}
    \vspace{-0.5cm}
    \caption{Comparison of the closed-loop performance of the proposed safety filter when using the class $\mathcal{K}$ decrease for NNs at different points in the training progress. The left plot shows the closed-loop trajectory of $y_{\mathrm{off}}$ for a single initial condition, with shaded constraint set. The right plot shows the distance of states with respect to the constraints $\mathcal{X}$ at time step $k=2000$ for $1000$ different initial states sampled from $\Dpb$, with respect to their $\ell_1$-loss over the training data set, denoted as $e_{\mathrm{\ell_1}}$. The error bars represent the standard deviations, with mean distances marked in black.}
    \label{fig:nonlinear_l1_multi_trajectory}
    \vspace{-0.3cm}
\end{figure}

\begin{figure}[t]
    \centering
    \vspace{0.3cm}
    \includegraphics[width=\linewidth]{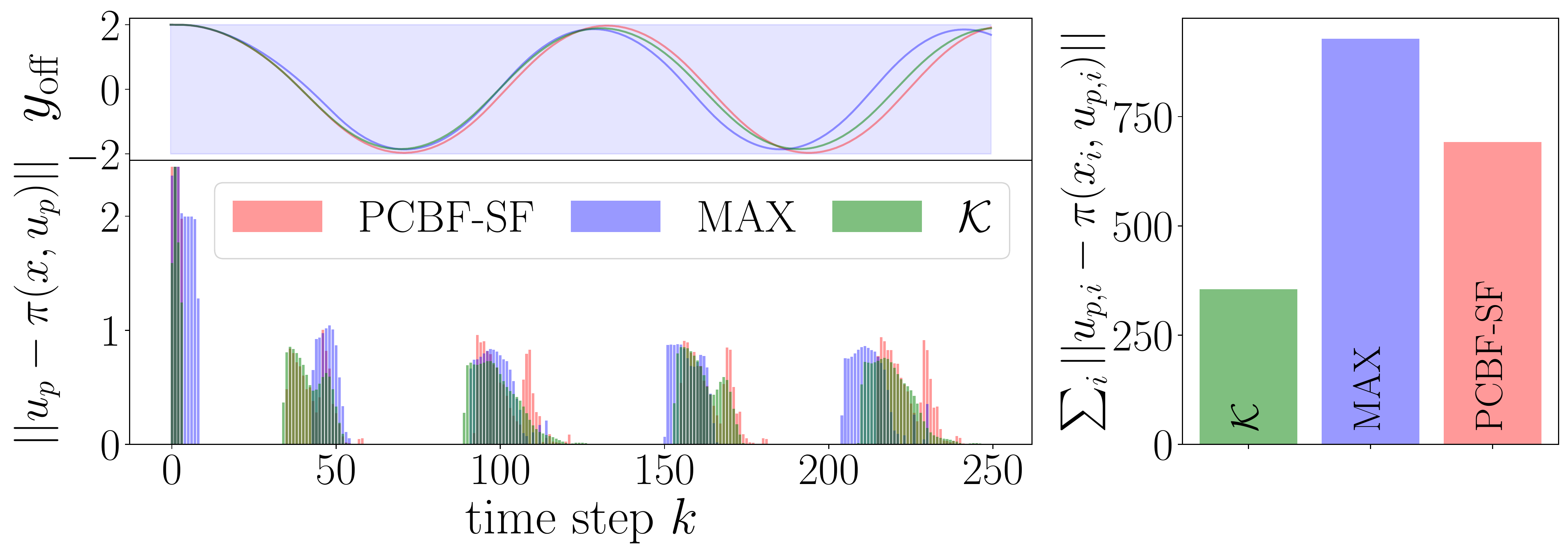}
    \vspace{-0.5cm}
    \caption{Comparison of the closed-loop performance of the proposed safety filter with maximum and hand-tuned class $\mathcal{K}$ decrease with the PCBF-SF in \cite{wabersich22}. The closed-loop position for one initial condition is shown at the top left and the amplitude of the respective safety interventions in the bottom left. The amplitude of safety interventions for one time step for 10000 randomly sampled states is shown on the right.}
    \label{fig:inputchange}
    \vspace{-0.3cm}
\end{figure}

\paragraph*{Simulation Results} We test our approach for the $\Delta \hpb$ approximation using the 
maximum CBF decrease function in \eqref{eq:approx_cbf_max_decrease} as well as a hand-tuned class $\mathcal{K}$ function $\Delta\hat{h}_\mathrm{PB}(x) = 0.5 \cdot \approxhpb(x)$ similar to \cite{zeng21}, with tolerance parameter $\delta_{\mathrm{tol}}=10^{-6}$. Note that the hand-tuned decrease function no longer guarantees recursive feasibility of the scheme but shows a less conservative safety filter behavior and allows solving only one optimization problem, i.e., only solving \eqref{eq:nn_cbf_sf}.
The simulation for multiple initial conditions is shown in Figures~\ref{fig:nonlinear_pcbf_sf},~\ref{fig:nonlinear_maxdec_multi_ic}~and~\ref{fig:nonlinear_kinf_multi_ic} for the original PCBF-based approach in \cite{wabersich22}, the approximate safety filter using the maximum decrease and the class $\mathcal{K}$ decrease, respectively. All three approaches are able to steer the closed-loop system starting at $y_{\mathrm{off}}(0) \notin \mathcal{X}$ back into the state constraints, with the class~$\mathcal{K}$ approach requiring a less strict convergence. In Figure~\ref{fig:nonlinear_l1_multi_trajectory} the effect of the approximation error of the learned NN on the state evolution of the closed-loop system is shown for the class $\mathcal{K}$ decrease function. As illustrated in both plots, NNs with an on average smaller approximation error $e_h$, which we estimate using the $\ell_1$-loss over the training data set, result in state trajectories converging to a region closer to the state constraints $\mathcal{X}$, whereas for a large approximation error, the closed-loop system no longer converges to the state constraints. The closed-loop safety filter interventions can be observed in Figure~\ref{fig:inputchange} for one initial condition and for 10000 randomly sampled states, where the approximate safety filter with maximum decrease shows the highest amplitude of interventions and the class $\mathcal{K}$ decrease uses the lowest.

For the solve time measurements, we sample $10^4$ states inside $\Dpb$ and state the results in Table~\ref{tab:nonlinear_time}. We observe a $34$-fold decrease in minimum solve time compared to the original PCBF-based algorithm and a $58$-fold decrease in average solve time. In general, larger neural network architectures lead to a larger solve time. When we double the number of parameters (factor of $1.9$, from 2 hidden layers to 3 hidden layers with 64 neurons), the minimum solve time increases by $17 \% $ for the approximate safety filter using the class $\mathcal{K}$ function decrease. As expected, the approximate safety filter using the maximum decrease takes on average $45 \%$ longer to solve than the method with the class $\mathcal{K}$ function decrease.
\begin{table}[h!]
\centering
\caption{Solve Time Measurements}
\vspace{-0.1cm}
\label{tab:nonlinear_time}
\renewcommand{\arraystretch}{1.3}
\begin{tabular}{|c | c | c | c|}
\hline
\textit{Nonlinear System} & \multicolumn{3}{c|}{\textbf{Solve Time [ms]}} \\
\hline
$\Delta\hat{h}_\mathrm{PB}(x)$ (\textit{NN Architecture}) &\textit{Minimum} & \textit{Average} & \textit{Std. Dev}. \\
\hline
Maximum Decrease $(2 \times64)$ & 2.551 & 4.571 & 1.234  \\
Maximum Decrease $(3 \times64)$  & 2.920 & 5.257 & 1.458 \\
\hline
Class $\mathcal{K}$ function  $(2 \times64)$ & \textbf{2.129} & \textbf{3.040} & \textbf{0.465 }  \\
Class $\mathcal{K}$ function $(3 \times64)$ & 2.491 & 3.495  & 0.580   \\
\hline
\hline
PCBF safety filter \cite{wabersich22} &  72.51 & 176.5 & 197.3 \\
\hline
\end{tabular}
\end{table}


\section{CONCLUSION}
This work presented an approximate predictive control barrier function (PCBF) scheme. The proposed approach relies on a neural network approximation of the predictive CBF presented in~\cite{wabersich22}, reducing the problem to one optimization problem. We present convergence guarantees to a neighbourhood of the safe set defined by the exact PCBF method, which depends on the approximation error and relies on a novel class $\mathcal{K}$ lower bound result on compact sets. 
We show in an example how the proposed method can be applied to recover an autonomous car outside the state constraints in simulation using different design methods and how the approximate method leads to a significant decrease in computation times.

\section*{DATA AVAILABILITY STATEMENT}
\vspace{-0.1cm}
The code and data in this study are available in the ETH Research Collection: \\ \href{https://doi.org/10.3929/ethz-b-000610171}{https://doi.org/10.3929/ethz-b-000610171}.


\bibliographystyle{IEEEtran}
\bibliography{root}

\end{document}